\documentclass[pra,twocolumn,showpacs,preprintnumbers,amsmath,amssymb]{revtex4}

\usepackage{graphicx}
\usepackage{dcolumn}
\usepackage{bm}
\usepackage{hyperref}
\newcommand{\hmin}{\ensuremath{H_\infty}}
\newcommand{\hminT}{\widetilde{H}_\infty}
\begin{document}

\preprint{APS/123-QED}
\title{Secure self-calibrating quantum random bit generator}

\author{M.\ Fiorentino}
 \email{marco.fiorentino@hp.com}
 \affiliation{Hewlett-Packard Laboratories,
1501 Page Mill Rd.\, MS 1123, Palo Alto, CA 94304-1100, USA}

\author{C.\ Santori}
 \affiliation{Hewlett-Packard Laboratories,
1501 Page Mill Rd.\, MS 1123, Palo Alto, CA 94304-1100, USA}

\author{S.\ M.\ Spillane}
 \affiliation{Hewlett-Packard Laboratories,
1501 Page Mill Rd.\, MS 1123, Palo Alto, CA 94304-1100, USA}

\author{W.\ J.\ Munro}
 \affiliation{Hewlett-Packard Laboratories, Filton Road, Stoke Gifford,
 Bristol BS34 8QZ, United Kingdom}

\author{R.\ G.\ Beausoleil}
 \affiliation{Hewlett-Packard Laboratories,
1501 Page Mill Rd.\, MS 1123, Palo Alto, CA 94304-1100, USA}

\date{\today}

\begin{abstract}
random-bit generators (RBGs) are key components of a variety of
information processing applications ranging from simulations to
cryptography. In particular, cryptographic systems require
``strong'' RBGs that produce high-entropy bit sequences, but
traditional software pseudo-RBGs have very low entropy content and
therefore are relatively weak for cryptography. Hardware RBGs
yield entropy from chaotic or quantum physical systems and
therefore are expected to exhibit high entropy, but in current
implementations their exact entropy content is unknown. Here we
report a quantum random-bit generator (QRBG) that harvests entropy
by measuring single-photon and entangled two-photon polarization
states. We introduce and implement a quantum tomographic method to
measure a lower bound on the ``min-entropy'' of the system, and we
employ this value to distill a truly random-bit sequence. This
approach is secure: even if an attacker takes control of the
source of optical states, a secure random sequence can be
distilled.
\end{abstract}

\maketitle

Random numbers are commonly used in computer simulations,
lotteries, and, most importantly, cryptographic applications.
Cryptographically strong random numbers need to have two
properties: good statistical behavior and unpredictability. The
numbers need to be distributed according to a unform distribution,
and an attacker should not be able to predict the corresponding
sequence of bits. Unpredictability is quantified using the entropy
content of a sequence generated by a random-bit generator (RBG)
\footnote{It has been shown (see e.g. Ref.\ \cite{ANSI}) that
min-entropy is the entropic quantity to be used to characterize
RBG. We will define min-entropy rigorously in this paper and show
how it can be measured in a particular system, in this
introduction we will use entropy content in a loose sense as a
measure of the randomness.}.

The entropy content can be used to grade RBG security, i.e., the
ability of the generator to withstand attacks.  Most applications
generate long strings of bits using algorithms known as
pseudo-random number generators, with seeds chosen by the user.
The entropy content of the strings generated in this fashion is
small and is ultimately determined by the length of the (short)
seed. This deficiency makes pseudo-random numbers unsuitable for
the most demanding cryptographic applications. This fact has been
recognized by both the information theory community and the
computer security industry \cite{ANSI,RFC}. Hardware RBGs are an
alternative to pseudo-RBGs because they harvest and distill
entropy from physical systems. The most recent examples of
hardware RBGs stress the importance of directly measuring the
entropy content of the source \cite{turbid}.

In principle, random bits could be produced by classical physical
processes that are too complicated to predict perfectly over long
times, such as thermal noise.  For example, Denker has used
thermal noise fluctuations in a resistor as a randomness source,
and relied on an \emph{estimate} of the entropy of the noise
process to extract a random bit sequence from digits derived from
that source \cite{turbid}. Further, sufficiently powerful data
processing systems with appropriate models or algorithms may
become able to predict chaotic or thermal processes, even if only
for a short time.

In quantum phenomena the outcome of a class of measurements is
governed by probabilistic laws: the statistical properties of
repeated measurements can be predicted, but the result of each
measurement is random.  This irreducible randomness of the quantum
phenomena is postulated here and is the basis of our RBG.
Distinguishing between irreducible quantum randomness and
classical randomness, that can in principle be controlled and
influenced, is at the basis of our RBG security.

Quantum measurements can be easily used to generate random bits.
For example, if we detect the transmission and reflection of a
$45^\circ$-polarized photon (a ``qubit'') on an
horizontal-vertical (H-V) polarizing beam-splitter with two
photomultipliers, each detector has the same probability to
register an event, but at any given time we cannot predict which
detector will record the next event. By assigning the value 0 to a
detection in one of the detectors and 1 to the other we can build
sequences of random numbers. Similarly, we can use pairs of
polarization-entangled photons that are described by
\begin{equation}
|\psi_+\rangle = \frac{|H_1 V_2\rangle + |V_1
H_2\rangle}{\sqrt{2}} , \label{one}
\end{equation}
so that appropriately balanced coincidence measurements in the
$H_1$-$V_2$ and $V_1$-$H_2$ basis yield equiprobable outcomes.
This type of quantum coin tossing has already been exploited for
the generation of random bits \cite{id,rnd1,rnd2}.  None of those
quantum RBGs presented a security analysis or a method to verify
integrity.

In this work we demonstrate a \emph{quantum} random-bit generator
(QRBG) based on measurements made on quantum states that span a
$2\times2$ Hilbert (sub)space. While there are a number of quantum
systems that could readily satisfy this constraint, we have
emphasized an optical implementation because of the ease with
which quantum states can be generated and measured. We follow
recent work on entropic statistical analysis of random sources
\cite{extractors1,extractors2,mine}, and we measure a quantity
known as ``min-entropy,'' $\hmin$, and use the value of $\hmin$ to
distill a random sequence of bits from a series of detection
events using a hash function.

Our approach has two main advantages over existing QRBGs. First,
we are able to measure and monitor continuously the randomness of
the bits, relying on a \emph{physical} property of the system.  We
do not rely on \emph{a-posteriori} statistical tests of generated
bit sequences, because these tests cannot prove randomness unless
they analyze infinite sequences.  Second, using this protocol
allows us to endow an attacker with more capabilities than any
other RBG: even if she takes complete control of the source of
optical states, so long as $\hmin > 0$ a sequence of bits
nevertheless can be extracted that is arbitrarily close to a
string of bits that is perfectly random. \cite{mine}

To define and measure the security of a RBG we must define the
adversarial context in which it operates.  In such scheme one has
to assume that the attacker has complete knowledge of the protocol
used and can, in principle, control or influence part of it. This
is similar to the scenarios used for quantum key distribution in
which the attacker has complete control of the communication
channels and knowledge of the protocol but has no access to the
transmission stations.

In our scenario, the user (Alice) can choose the quantum system on
which she makes a measurement to generate random bits but the
adversary (Eve) controls the state of the quantum system but has
no access to the measurement apparatus (the tomography setup, in
our case). Notice that Alice is not allowed to exploit other
degrees of freedom different from the ones under Eve's control.
This restriction is due to the fact that one must assume an
attacker has knowledge of the protocol and will try to gain
control of the degrees of freedom that are actually being used for
generating the random numbers. Even using such unfavorable
scenario for Alice we demonstrate that a secure RBG can be built
using such assumptions. This is a worst-case scenario: our
protocol is secure \emph{a fortiori} if Eve has less than total
control of the state of the system or if she tries to exploit
failures in the system to gain knowledge of the random bits.

One could argue that our adversarial scenario is somewhat
contrived because Eve is not likely to gain control of the source.
There are two arguments to counteract such objection. First,
protocol robustness is increased if one shows that it is resilient
against a larger class of attacks.  Second even if Eve does not
control directly the degree of freedom used to generate the random
numbers she can nevertheless take advantage of a system failure to
gain knowledge of the bits being generated.  In this respect our
protocol is more secure than any other hardware random number
generator we know of.

In our protocol Alice picks the simplest quantum system, a qubit,
and makes a projective measurement to generate random bits. In
this contest, we believe, simplicity is a virtue and this is the
reason for using a qubit. This allows a complete analysis and
excludes the possibilities of extra degrees of freedom used as
"back-doors" by Eve.  More complicated systems might have similar
security but are outside the scope of this paper.

Here we implement the qubit in the polarization of photons. The
polarization state of the photons is controlled by Eve, but she
has no knowledge of the sequence of measurements made by Alice
except for the basis used for the projection measurement used to
generate the random bits.\footnote{This adversarial scheme
includes the case in which Eve uses an entangled state to prepare
remotely Alice's state. This more complicated preparation scheme
does not give Eve any advantage.} For any other hardware RBG one
requires that Eve has no control over the randomness source while
in our adversarial scenario she completely controls one component
(i.e. state preparation) of the source.

Alice's measurement strategy is consistent with the provision of a
$2\times2$ Hilbert space (i.e. a qubit), and that any state Eve
sends to Alice can be represented by a $2\times2$ complex density
matrix $\hat{\rho}$. For any density matrix $\hat{\rho}$, Eve can
try to bias the output of the QRBG in a way that is known to her,
but appears random to Alice, by sending a collection of pure
states $|\psi_i\rangle$ with corresponding probabilities $p_i$
such that
\begin{equation}
\hat{\rho}=\sum_i p_i |\psi_i \rangle \langle \psi_i| \, ;
\label{decomposition}
\end{equation}
i.e., she can use any decomposition of $\hat{\rho}$. Eve cannot
control the outcome of a measurement on the pure state
$|\psi_i\rangle$ (because these probabilities are governed solely
by the laws of quantum mechanics), but knows at each time the
state Alice is measuring.  How much information can Eve obtain in
this case about Alice's random sequence?  Or, in other words, how
can Alice separate the quantum randomness from the classical one?

We begin to answer these questions by defining an entropic
quantity known as the min-entropy \cite{mine}:

\textbf{Definition 1} The min-entropy of a random variable $X$,
 denoted by $\hmin(X)$, is
\begin{equation}
\hmin(X) \equiv -\log_2(\max_{x \in X}\textrm{P}(x))
\end{equation}
where $\textrm{P}(x)$ is the probability of a particular outcome
of the random variable $X$. For a secure implementation the
probabilities $\textrm{P}(x)$ should be calculated from the
attacker point of view and a worst-case scenario regarding the
amount of her knowledge. When so defined the min-entropy can be
used to determine the quality of a source of randomness. For a
binary variable, $\hmin = 1$ corresponds to a completely random
process, and $\hmin = 0$ to a deterministic one.

Alice generates $n$ bits by measuring the states provided by Eve.
If the bits were generated by measuring $n$ times a qubit in the
pure state $|\psi\rangle$ in the computational basis $|0\rangle,
|1\rangle$, then the min-entropy will be
\begin{eqnarray}\nonumber
\hmin(|\psi \rangle \langle \psi|^n)&=&- n \log_2
(\max(|\langle 0|\psi\rangle|^2,|\langle 1|\psi\rangle|^2)\\
 &\equiv&- n \log_2 (\max(\textrm{P}_{0},\textrm{P}_{1})). \label{pure}
\end{eqnarray}
This definition can be extended to a decomposition such as the one
on the RHS of Eq.\ \ref{decomposition},
\begin{eqnarray}
& &\hmin\left[\left(\sum_i p_i |\psi_i \rangle \langle
\psi_i|\right)^n\right] \label{def}\\
\nonumber &=&- n  \sum_i p_i
\log_2(\max(P_{0}(|\psi_i\rangle),P_{1}(|\psi_i\rangle)))\\
\nonumber&=& n  \sum_i p_i \hmin \left(|\psi_i \rangle \langle
\psi_i|\right) \, .
\end{eqnarray}

Since Alice does not know anything about the decomposition that
Eve may be using, we will define the min-entropy of a state
$\hat{\rho}$ (denoted $\hminT(\hat{\rho})$) to be the minimum
value of the min-entropy taken over \emph{all possible}
decompositions of $\hat{\rho}$. This approach allows us to put an
upper bound on the amount of information Eve can obtain about
Alice's sequence, and to determine the worst-case parameters for
the randomness extractor that is used below. \cite{mine}

By assumption, $\hat{\rho}$ is a $2 \times 2$ density matrix, so
that without loss of generality we can write
\begin{eqnarray}
 \hat{\rho}(S_1,S_2,S_3)
  = \frac{1}{2}\left(%
\begin{array}{cc}
  1+S_3 & S_1- i S_2 \\
  S_1+ i S_2 & 1-S_3 \\
\end{array}%
\right) \label{dm}
\end{eqnarray}
where $S_{1,2,3}$ are the real Stokes parameters (for $S_0 = 1$)
for the qubit space. The point $(S_1,S_2,S_3)$ lies inside or on
the Poincar\'{e} sphere for physical density matrices.

\textbf{Definition 2} We define the function $f(\hat \rho)$, which
is real valued for all physical density matrices, as
\begin{equation}
f(\hat \rho) = -\log_2\left(\frac{1 + \sqrt{1- |S_1- i
S_2|^2}}{2}\right) . \label{effe}
\end{equation}

We can now state the theorem that is the centerpiece of our QRBG
algorithm:

\textbf{Theorem} The min-entropy of a system described by an
arbitrary density matrix $\hat \rho$ is
\begin{equation}
\hminT(\hat{\rho}) = f(\hat \rho). \label{th}
\end{equation}
This theorem can be demonstrated using the following three lemmas,
which are easily established:\cite{demo}

\textbf{Lemma 1} For each pure state $|\psi\rangle$
\begin{equation}
\hmin\left(|\psi \rangle \langle \psi|\right) = f\left(|\psi
\rangle \langle \psi|\right).
\end{equation}

\textbf{Lemma 2} The two pure states represented by the density
matrices
\begin{equation}
|\eta_\pm\rangle\langle\eta_\pm| =\frac{1}{2}\left(%
\begin{array}{cc}
  1 \pm S_3' &  S_1- i S_2\\
  S_1+ i S_2 & 1 \mp S_3' \\
\end{array}%
\right) \label{dc}
\end{equation}
with $S_3'=\sqrt{1-S_1^2-S_2^2}$, are a valid decomposition of the
density matrix in Eq.\ \ref{dm}.

\textbf{Lemma 3} The function $f\left[\hat \rho\left(S_1,S_2,S_3
 \right) \right]$ is a convex function of $S_1$, $S_2$, and $S_3$
 in the Poincar\'{e} sphere.

Using the convexity of $f$ we can write
\begin{equation}
f(\hat{\rho}) \leq \sum_i p_i f\left(|\psi_i \rangle \langle
\psi_i|\right) \label{convex}
\end{equation}
for each decomposition of $\hat{\rho}$. Using Eq.\ \ref{def} and
the result of Lemma 1 we obtain
\begin{equation}
f(\hat{\rho}) \leq \hmin \left(\sum_i p_i |\psi_i \rangle \langle
\psi_i|\right) \label{final}
\end{equation}
indicating that $f\left(\hat\rho\right)$ is a lower bound for
$\hminT (\hat\rho)$. Using Lemma 1, we can show that the
decomposition of Lemma 2 has a min-entropy equal to
$f\left(\hat\rho\right)$, and therefore that
$f\left(\hat\rho\right)$ is equal to the minimum of $\hmin$ over
all possible decompositions of $\hat \rho$, i.e.,
$f\left(\hat\rho\right) = \hminT\left(\hat\rho\right)$. From this
demonstration, it follows that the decomposition of Lemma 2 is the
optimal choice for Eve, since it leads to the most pessimistic
estimate of the min-entropy of the source.

The theorem provides a link between the density matrix and the
source min-entropy. The latter quantity is interesting because of
the vast computer science literature on entropy extractors (see,
e.g., the review papers \cite{extractors1,extractors2}). An
entropy extractor---such as the example given by Ref.\ \cite{mine}
used in our work here---is an algorithm that accepts an imperfect
source of random bits and outputs a sequence arbitrarily close to
a uniformly distributed sequence \footnote{The impossibility proof
of Ref.\ \cite{imposs} does not apply in this case because we are
using a probabilistic extractor.}. Given a raw $n$-bit sequence
the algorithm allows one to extract an $m$-bit privacy-enhanced
sequence which is arbitrarily close to a uniform distribution,
where
\begin{equation}
m = \hminT n - 4 \log_2(1/\epsilon) - 2 \label{m}
\end{equation}
and $\epsilon$ is the statistical distance between the
distribution of the $m$ bits and a uniform distribution. We refer
the reader to Ref.\ \cite{mine} for a proof of the security of the
extraction algorithm, and to Ref.\ \cite{demo} for the technical
details of the particular algorithm we implemented.
\begin{center}
\begin{figure*}
\includegraphics[width=7in]{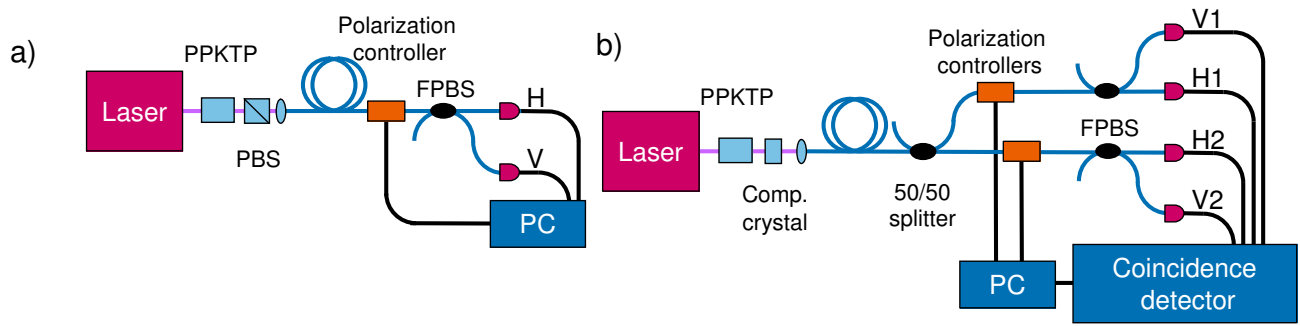}
\caption{Schematics of the QRBGs using single-photons (a) and
entangled pairs (b). PPKTP is the nonlinear crystal, PBS the bulk
polarization beam-splitter, FPBS the fiber polarization
beam-splitter.\label{fig}}
\end{figure*}
\end{center}
\begin{figure}
\begin{center}
\includegraphics[width=2.3in]{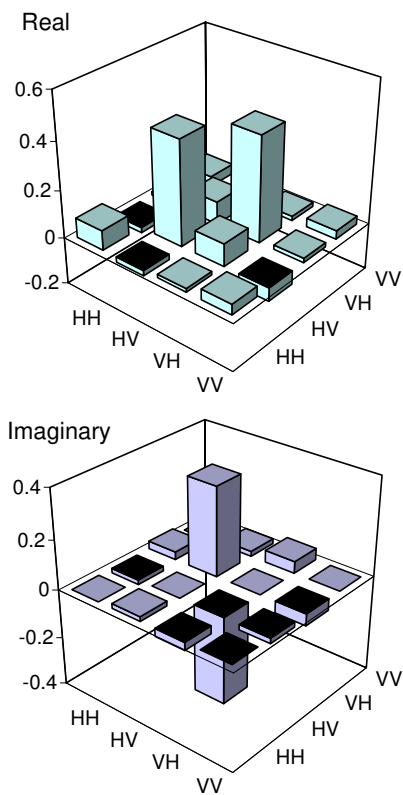}
\caption{Real and imaginary part of the density matrix for the
photon pair polarization state used to generate the random-bit
sequence.\label{densmat2}}
\end{center}
\end{figure}
We realized the two implementations of the QRBG shown in Fig.\
\ref{fig}. The first implementation [Fig.\ \ref{fig}(a)] uses a
linearly polarized source with average intensity at the
single-photon level. We used photons extracted from pairs
generated by spontaneous parametric down-conversion (SPDC) in a
periodically-poled potassium titanyl phosphate (PPKTP) crystal;
however, either an attenuated laser or LED could have been used
instead. We used parametric down-conversion in a 10-mm crystal
manufactured by Raicol Crystals with a poling period of $10 \mu$m.
In the crystal a photon from the violet laser diode (13 mW at a
wavelength of 405 nm, Sacher Lasertechnik, TEC-100-405-20) is
split into a pair of orthogonally polarized infrared photons with
a wavelength of 810 nm and a 1-nm bandwidth defined by an
interference filter. The photons are coupled into a single-mode
fiber, propagate through a polarization-controlling stage and are
split in two approximately equal parts on a fiber polarization
beam-splitter. The photons are recorded by photodetectors, and
each detection event is recoded as a random bit (0 for
horizontally polarized photons, and 1 for vertically polarized
photons). The photons' density matrix is tomographically
reconstructed off-line \cite{tomo}.  Using the density matrix and
our theorem, we compute the min-entropy $\hminT = 0.96$, and we
input this value to the randomness extractor \cite{mine}. The
raw-bit generation rate is 60 kbits/s, and the bits are passed to
the randomness extractor to obtain a bit-generation rate of
approximately 57 kbits/s. A sample file containing 100 million
random bits thus obtained is available online \cite{online}.

The second implementation [Fig.\ \ref{fig}(b)] uses
polarization-entangled photon pairs described by the state of Eq.\
\ref{one}. The entangled photons, generated by SPDC in the PPKTP
crystal followed by post-selection \cite{chris}, are sent to
polarization controllers, fiber polarization beam-splitters, and
single-photon detectors for analysis. Coincidence events are
recorded as random bits  (0 for $H_1$-$V_2$ and 1 for
$V_1$-$H_2$). By restricting the measurement to the coincidences,
we effectively restrict the 2-qubit space of the photon pair to a
2-dimensional Hilbert subspace described by an effective-qubit
state.  By carrying out a complete tomography of the 2-qubit state
\cite{tomo} we can extract the effective-qubit density matrix and
the relative min-entropy. Figure \ref{densmat2} shows a
reconstructed density matrix corresponding to a min-entropy of
$\hminT = 0.38$. Figure \ref{densmat2} shows that the fiber
birefringence changes the state without affecting the min-entropy,
and we do not subtract accidental coincidences from the
tomographic data. (Such a correction, in fact, would increase the
min-entropy, but weaken the security of the protocol). The raw bit
rate for this QRBG is 14 kbits/s, while the random-bits rate is
5.3 kbits/s. Again, a sample file with 100 million random bits is
available online \cite{online}.

We have applied a battery of \emph{a posteriori} software
statistical tests to the privacy enhanced output, but we stress
that these tests are only used to verify that the QRBG has been
correctly implemented: the guarantee of the QRBG security and
randomness relies on the measurement of $\hmin$. We used the NIST
test suite \cite{nist}, which consists of a set of 15 statistical
tests of random numbers for cryptographic applications. Our QRBGs
pass the test, and the detailed test results are given online
\cite{online}.

A comparison between the two implementations of the QRBG makes it
obvious that the single-photon implementation is simpler and has
much higher bit flux. The entangled-photon implementation has the
advantage that, by using coincidences, much of the stray-light
noise is suppressed. However, by carefully screening the detection
apparatus, the effect of stray photons can be made negligible even
in the single-photon case.

Let us review here the advantages of our quantum RBG when compared
with other implementations.  Compared with pseudo-random number
generators our hardware RNG has the advantage of generating bit
sequences with full entropy. Other hardware random number
generators are based on chaotic systems \cite{turbid} that can be,
in principle, predicted or influenced; our quantum RNG relies on
quantum measurements that are, as far as we know, fundamentally
random. In addition Ref.\ \cite{turbid} uses an estimate of the
Shannon entropy (not the min-entropy) that is realized once for
all: the user cannot continuously monitor the entropy to verify
the security and integrity of the RBG.  Compared with other
quantum RBGs \cite{id,rnd1,rnd2} our implementation is the first
that explicitly takes into account security.  To guarantee
security the min-entropy has to be \emph{measured} and filtering
has to be applied in a way that is analogous to the error
correction and privacy amplification routine used in quantum key
distribution protocols. References \cite{id} and \cite{rnd1} use
an experimental setup that is conceptually similar to the single
photon setup of Fig.\ \ref{fig} whereas the polarization beam
splitter is substituted with a non-polarizing 50/50 beam splitter.
For these implementations an attack scenario equivalent to the one
we have analyzed would involve giving Eve control over the beam
splitter. She could, for example, substitute the beam splitter
with a switch and therefore completely control the outcome of the
RBG.  To guarantee security and integrity of this kind of RBG
Alice needs to verify that the photons are coherently split among
the output arms of the beam splitter and that the coherence is
collapsed by her measurement. She can do so by making
interferometric measurements that are formally analogous to the
one we make but are more complicated from an experimental point of
view.  For these reasons we used the polarization scheme to
implement our secure RBG.

A number of improvements in our setup are possible.  The raw bit
rate is currently limited by the data acquisition hardware, so
dedicated hardware can speed up the acquisition and eliminate this
bottleneck. Eventually the bit rate will be limited by the dead
time in the detectors. Based on a comparison with existing QRBGs
\cite{id} we expect that rates up to several Mbits/s can be
achieved. Using off-line tomography relies on the assumption that
the system state does not change in the interval between the
measurement of $\hminT$ and the acquisition of the random bits.
While this is the case in the current implementation, on-line
tomography will both relax this assumption and increase the bit
rate. We are currently engineering a high-performance system in
which on-line tomography is carried on at the same time as the raw
bits are acquired.  We also observe that at this point the
security of the protocol is limited to individual attacks; further
analysis is needed to extend the security proof to attacks in
which Eve sends Alice clusters of entangled photons.

In conclusion, we have defined the worst-case min-entropy of a
qubit and introduced a method to measure it using quantum
tomography. Based on the properties of the min-entropy, we
constructed two implementations of a self-calibrating random
number generator which is secure against a large class of attacks.
We believe that our RBG will have important technological impact
in the area of secure communications and that, properly extended,
the min-entropy defined here could prove to be an important tool
in defining the security of qubit-based communication protocols.

This work was supported by DARPA through seed program number
HR0011-04-3-0040.

\end{document}